\documentstyle[12pt,epsfig]{article}
        \newdimen\eqskip
        \newdimen\txtskip
        \baselineskip=\txtskip
\begin{document}

  \newcommand{\ccaption}[2]{
    \begin{center}
    \parbox{0.85\textwidth}{
      \caption[#1]{\small{{#2}}}
      }
    \end{center}
    }
\def    \be             {\begin{equation}}
\def    \ee             {\end{equation}}
\def    \ba             {\begin{eqnarray}}
\def    \ea             {\end{eqnarray}}
\def    \nn             {\nonumber}
\def    \=              {\;=\;}
\def    \frac           #1#2{{#1 \over #2}}
\def    \ret            {\\[\eqskip]}
\def    \ie             {{\em i.e.\/} }
\def    \eg             {{\em e.g.\/} }
\def    \lsim           {\raisebox{-3pt}{$\>\stackrel{<}{\scriptstyle\sim}\>$}}
\def    \bentarrow      {\:\raisebox{1.1ex}{\rlap{$\vert$}}\!\rightarrow}
\def    \rd             {{\mathrm d}}    
\def    \Im             {{\mathrm{Im}}}  
\def    \bra#1          {\mbox{$\langle #1 |$}}
\def    \ket#1          {\mbox{$| #1 \rangle$}}
\def\mbc{m_{B_c}} 
\def\mbu{m_{B_u}} 
\def\fbc{f_{B_c}} 
\def\fbu{f_{B_u}} 
\def\bc{B_c}   
\def\bu{B_u}
\def\taubc{\tau_{B_c}}
\def\taubu{\tau_{B_u}}
\def\fbtobc{f(b\to \bc)}
\def\fbtobu{f(b\to \bu)}
\def\vcb{V_{cb}}
\def\vub{V_{ub}}

\def    \kev            {\mbox{$\mathrm{keV}$}}
\def    \mev            {\mbox{$\mathrm{MeV}$}}
\def    \gev            {\mbox{$\mathrm{GeV}$}}

\begin{titlepage}
\nopagebreak
{\flushright{
        \begin{minipage}{5cm}
        CERN-TH/97-150\\
        hep-ph/9707248\\
        \end{minipage}        }

}
\vfill
\begin{center}
{\LARGE { \bf \sc The Contribution of ${\rm B_c}$ Mesons
    to \\[1cm]       
     the Search for $B^+ \to \tau^+ \nu_{\tau}$ Decays at LEP}}
\vfill                                                         
\vskip .5cm
{\bf Michelangelo L. MANGANO\footnote{On leave of absence from 
    INFN, Pisa, Italy}}\\
{CERN, TH Division, Geneva, Switzerland} \\
\verb+ mlm@vxcern.cern.ch+\\
\vskip .5cm
{\bf S.R.~SLABOSPITSKY}\\
{Institute for High Energy Physics, Protvino, Moscow Region,
142284 Russia}\\
\verb+ slabospitsky@mx.ihep.su+\\
\end{center}                  
\nopagebreak
\vfill
\vskip 3cm
\begin{abstract} 
We study the contribution of $B_c$ mesons  to the search for  $B\to\tau
\nu_{\tau}$ decays.  We find that at LEP the contributions from $B_u$ and $B_c$
mesons can be comparable. This observation can have a relevant impact on the
extraction of constraints on new physics (such as charged-Higgs contributions) 
from current LEP limits on $B\to \tau\nu$ final states. Inclusion of the $\bc$
contribution can reduce the current L3 limit
on $\tan\beta/M_H$ from $0.38$~GeV$^{-1}$ (90\%CL) down to
$0.27$~GeV$^{-1}$ (90\%CL).
\end{abstract}                                                          
\vskip 1cm
CERN-TH/97-150\hfill \\
\today \hfill 
\vfill 
\end{titlepage}
The study of purely leptonic decays of heavy charged mesons $P^+$ (e.g. 
$D^{\pm}$,  $D^{\pm}_s$, $B^{\pm}$, ...)                                
\begin{eqnarray}
P^+(Q \bar q) \, \to \, \ell^+ \nu_{\ell}    \label{eq1} 
\end{eqnarray}                           
is of particular interest, due to their sensitivity to both the meson decay
constants~$f_P$ and the CKM matrix elements $V_{qQ}$. In the Standard Model
(SM)  the width of the decay (\ref{eq1}) is predicted to be  (here we assume a
zero value for neutrino mass):              
\begin{eqnarray}                               
\Gamma_{SM}(P^+ \to \ell^+ \nu_{\ell} ) = \frac{G_F^2 }{8 \pi} |V_{Qq}|^2
 f^2_{P} M_P m^2_{\ell} (1 - \frac{m^2_{\ell}}{M^2_P})^2 \; , \label{eq2}
\end{eqnarray}                                           
where $G_F$ is the Fermi coupling constant and
$M_P$ and $m_{\ell}$ are the masses of the pseudoscalar meson $P$ and
of the charged lepton $\ell^+$, respectively.

The widths of such decays are also very sensitive to the possible presence of 
physics beyond the SM.
For example, in models with two Higgs doublets the decay width
given in eq.~(\ref{eq2}) acquires the additional factor due the annihilation
via a charged Higgs ($H^{\pm}$)~\cite{ch}:
\begin{eqnarray}                
\Gamma_{H^{\pm}}(P^+ \to \ell^+ \nu_{\ell} ) = \Gamma_{SM}(P^+ \to \ell^+
\nu_{\ell} ) 
\times \left ( 1 - \frac{\tan^2\beta}{M^2_H} M^2_P \right )^2, \label{eq3}
\end{eqnarray}
where $\tan \beta$ is the ratio of the vacuum expectation values for the Higgs
fields and $M_H$ is the mass of the charged Higgs boson.
Similarly, a possible admixture of a right-handed
$(V+A)$ current to the standard $(V-A)$ current could also lead to
modifications of eq.~(\ref{eq2}) (see \cite{rc1, rc2}).     
                                          
Because of the strong helicity suppression, $e$ and $\mu$ leptonic
decays of $B$ mesons are very far from being observed. 
On the contrary, the statistics available
today at CLEO and LEP or becoming available in the very near future at the
$B$-factories should allow the first observation of
the $B\to \tau \nu_{\tau}$ decay mode. 
Notice that even in this case, however, the
$B_u\to \tau \nu_{\tau}$ decay has a very small 
branching ratio, of the order of few$\times 10^{-5}$. 
The current results of these searches       
at low energies (in the reaction 
$\Upsilon (4S) \to B^+ B^-$ \cite{argus, cleo}) as well as at the energy of the
LEP collider~\cite{l3, aleph} are collected in Table~1. 
The recent measurement by L3~\cite{l3},                           
based on a fraction of the full data sample accumulated during the LEP running,
set an upper limit on the  $B\to \tau \nu_{\tau}$ branching ratio (BR) which is
within less than an order of magnitude from the SM estimate. This limit is
nevertheless already sufficient to set interesting constraints on the allowed
values of $\tan\beta/M_H$~\cite{l3}~\footnote{In a recent paper~\cite{Soni},
the prospects for the extraction of tight constraints
on this parameters using $B\to D\tau\nu$ decays have also been considered.}
                                    
\begin{table}
\begin{center}
\vspace {0.5cm}
\begin{tabular}{|c||c|c|c|} 
\hline 
 Exp. & energy & ${\rm Br}(B \to \tau \nu)$ & 
 $\frac{\tan \beta}{M_H}, \,\, {\rm GeV}^{-1}$ \\ \hline
 ARGUS~\cite{argus} & $\Upsilon (4S)$ & $< \, 1.04 \times 10^{-2}$ 
& $ < 0.69$ \\ \hline
 CLEO~\cite{cleo} & $\Upsilon (4S)$ & $< \, 2.2 \times 10^{-3}$ 
& $ < 0.489$ \\ \hline
 ALEPH~\cite{aleph} & $Z$--boson & $< \, 1.8 \times 10^{-3}$ 
& $ < 0.468$ \\ \hline
 L3~\cite{l3} & $Z$--boson & $< \, 5.7 \times 10^{-4}$ 
& $ < 0.38$ \\ \hline
\end{tabular}
\ccaption{}{ Experimental upper limits for the branching ratio of
 $B^{\pm} \to \tau^{\pm} \nu_{\tau}$ decay and the evaluated constraint for
 charged Higgs boson parameters. }
\end{center}                      
\end{table}
   
In this note we point out that for measurements carried out
at energies well above the threshold for the production of $b\bar b$ pairs,
production and decay of $B_c$ mesons can give a substantial        
contribution to $\tau \nu_{\tau}$ final states~\footnote{This observation has
certainly been made already in the past. We recently learned, for example, 
that M. Neubert made this remark in several occasions. 
To our knowledge, however, a detailed study of its consequences, in the light
of the current measurements from LEP, has never been documented.}
This observation strengthens the current limits from LEP on $BR(B_u\to      
\tau\nu)$ and, as a result, on the possible presence of new physics.
As we shall see, the rate for $Z^0\to B_c X \to \tau\nu X$ could well be larger
than that for $Z^0\to B_u X \to \tau\nu X$, making this process a possible 
discovery channel for the as yet unobserved $B_c$ meson.

Bound states of a $b$ and $\bar c$ quark pair (the 
$B_c$ meson) have never been observed. They represent interesting objects
because their production and decay properties are expected to be calculable 
within QCD. Typical non-perturbative parameters such as the decay constant
$\fbc$ can be calculated by using, for example,
potential models or QCD sum rules (see for
example the review in ref.~\cite{gers} and the references therein). 
                          
The relative fraction of $\tau\nu$ final states coming from $\bc$ and $\bu$
production is given by:
\be  \label{eq:ratio}
    \frac{N_{\bc}}{N_{\bu}} \; = \;
    \frac{\fbtobc}{\fbtobu} \; \left\vert \frac{\vcb}{\vub}\right\vert^2 \;
    \left( \frac{\fbc}{\fbu}\right)^2 \; \frac{\mbc}{\mbu} \;
    \frac{\taubc}{\taubu} 
    \frac{(1-\frac{m^2_{\tau}}{m_{B_c}^2})^2 }
         {(1-\frac{m^2_{\tau}}{m_{B_u}^2})^2 } \; ,
\ee
where $\taubc$ and $\taubu$ refer to the $\bc$ and $\bu$ lifetimes, 
and the factor $\fbtobc$ ($\fbtobu$) is the inclusive 
probability that a $b$ quark hadronizes into a $\bc$ ($\bu$) meson. These
factors
include transitions in which the $b$ quark hadronizes into excited $\bu$ and 
$\bc$ states, which then decay strongly into the stable scalar mesons.
While $\fbtobc$ is estimated to be a small number, of the order of 0.1\%, the
large ratio $(\vcb/\vub)$ can largely compensate for the suppressed $\bc$
production rate.

Before turning to the numerical estimate of the $\bc$ contribution to the
$\tau\nu$ rate, we discuss the values and potential correlations
of the parameters present in eq.~(\ref{eq:ratio}).
We start from the parameters relative to the $\bu$ system. Mass and lifetime
are known today with good accuracy~\cite{pdg}:
\be                                     
\mbu=5.2789\pm 0.0018~{\rm GeV} \; , \quad \taubu = (1.62\pm 0.06)\times
         10^{-12} {\rm s} \; .              
\ee                                        
The probability $\fbtobu$ is known at LEP with accuracy better than
10\%~\cite{pdg}:                  
\be
\fbtobu = 0.378\pm 0.022 \; .
\ee                          
The value of the decay constant $\fbu$ has never been measured, but 
estimates based on lattice calculations have achieved a good degree of
accuracy, and give~\cite{Martinelli96}:   
\be                                                             
\fbu=175\pm 25 ~{\rm MeV} \; .
\ee
Of course the value of $\fbu$ is exactly one of the parameters which one would
like to extract from the measurement of the leptonic $\bu$ decay. As we will
show, doing this with sufficient accuracy requires the use of a sample of $\bu$
mesons without a $\bc$ contamination.
Use of the central value for the above parameters, in addition to 
using the central value of $\vub=0.0033\pm 0.0008$, gives the following
\be
      BR(\bu\to \tau\nu_{\tau}) = 5.8 \times 10^{-5} \; .
\ee                                                      

The only parameter relative to the $\bc$ meson which is expected to be known
with good accuracy from the current calculations is the 
mass~\cite{gers, potential}:
\be                                                     
     \mbc=6.275\pm 0.040~{\rm GeV} \; .
\ee                    

A recent thorough analysis of the $\bc$ lifetime, performed within NRQCD
and
including non-relativistic corrections up to order $v^2$, 
has been carried out in ref.~\cite{ben}. The result is given by:
\be  \label{eq:buBR}
   \taubc = (0.55\pm 0.15)\times
         10^{-12} {\rm s} \; ,
\ee                          
where the quoted uncertainty is dominated by the limited knowledge of the 
charm and bottom quark masses, and to a lesser degree by the uncertainty in the
strange quark mass and in $\vcb$. Because of the almost complete cancellation 
between the weak-annihilation rate and the effect of
Pauli-interference diagrams, no         
significant uncertainty on the lifetime arises from the limited knowledge of
the decay constant $\fbc$~\cite{ben}. 
This has been calculated both in potential models~\cite{gers, potential}
and by use of sum-rules~\cite{sumrules}.                          
A complete review of these results and a
comprehensive bibliography can be found in ref.~\cite{gers}. From this review
we extract the following range of values for $\fbc$:
\be
    \fbc=450\pm 100~{\rm MeV} \; .
\ee

The transition probability $\fbtobc$ at LEP has been evaluated by several
authors~\cite{ftobc}.                       
It is in principle calculable from QCD once the value of the meson wave
function is known. The leading production mechanism is through the
emission of an off-shell gluon from the $b$ quark, with the gluon splitting
into a $c\bar c$ pair and the $b\bar c$ system binding into the $\bc$ or
excited states thereof. Several
uncertainties are present in this calculation. For example, 
NLO QCD corrections are not
known, and the result is very sensitive on the choice of renormalization scale.
Likewise, the wave functions of the ground state state and of the excited
states are not known with high accuracy, as discussed above.
We shall therefore leave $\fbtobc$ a
free parameter, varying it in the range suggested by the envelope of the
available calculations ($0.02\% < \fbtobc < 0.1\%$) and studying 
the dependence of our results on it.           
                   
The final parameter which is needed to determine the quantity in
eq.~(\ref{eq:ratio}) is the ratio $\vub/\vcb$. 
The value for this ratio found in the RPP~\cite{pdg} is:
\be
      \frac{\vub}{\vcb} =0.08 \pm 0.02 \; .
\ee

Using the central values of the parameters discussed so far, we obtain the
following result:                   
\be  \label{eq:ratioc}
    \frac{N_{\bc}}{N_{\bu}} \; = \; 1.2 \; \left[\frac{\fbtobc}{
           10^{-3}}\right]         \; .
\ee                                     
The upper and lower values which can be obtained by taking the extreme 
ranges quoted for the parameters $\fbc$ and $\taubc$ are:
\be  \label{eq:ratiol}                       
    \left(\frac{N_{\bc}}{N_{\bu}}\right)_{\rm min}
      \; = \; 0.52 \; \left[\frac{\fbtobc}{ 10^{-3}}\right]
\ee                                                        
and
\be  \label{eq:ratioh}
    \left(\frac{N_{\bc}}{N_{\bu}}\right)_{\rm max}
      \; = \; 2.3 \; \left[\frac{\fbtobc}{ 10^{-3}}\right] \; .
\ee                
As is seen from eq.~(\ref{eq:ratiol}), even  the minimum relative 
contribution due to $B_c$ meson can have a noticeable magnitude~($\sim 50 $\%).
Therefore, this effect should be taken into account when one considers
the results of the search and analysis of the $B_u \to \tau \nu$ decay.

To demonstrate the importance of this $B_c$ contribution let us analyze the
results of the L3~Collaboration, which reported the following 
upper limit for the  branching ration of the $B_u \to \tau^{\pm} \nu_{\tau}$
decay~\cite{l3}:
\begin{eqnarray}
{\rm Br}(B_u \to \tau \nu) \, < 5.7 \times 10^{-4} \quad (90\% \, \, CL) \; .
\label{eq12}                                                                 
\end{eqnarray}
This is almost a factor of 10 larger than the 
SM prediction in eq.~(\ref{eq:buBR}).
In order to assess the impact of the $\bc$ production and decay on this
measurement, we would need an estimate of the relative tagging efficiency for 
the secondary vertex of $\tau$-tracks from $\bc$ and from $\bc$ decays. Since
the expected lifetime of the $\bc$ is a factor $2\div 3$ smaller than the
lifetime of the $\bu$, we expect a possible loss in efficiency, and a dilution
of the $\bc$ contamination of the $\tau\nu$ signal. 
We notice that in previous
searches for $\bc$ production at LEP~\cite{opalBc,delphiBc,alephBc}, 
using $\bc\to
\psi \ell\nu$ or $\bc \to\psi\pi$ decays, the changes in
efficiency due varying $\taubc$ in the range $0.4 \div 1.4$ps have been 
estimated to be small, of the order of 20\%. In absence of a specific study for
the $\tau\nu$ final state, we shall assume for definiteness that the $\bc$ and
$\bu$ decays have the same efficiency.

Using the eq.~(\ref{eq3}) and the result~(\ref{eq12}) the authors of 
\cite{l3} obtain the
following constraint for the contribution due to charged Higgs boson:
\begin{eqnarray}
\frac{\tan \beta}{M_{H}} (B_u) < 0.38 \quad \cite{l3}. \label{l30} 
\end{eqnarray}
Taking into account the $B_c$ meson contribution we can improve this
constraint. Our results are shown in Table~2, for several values of $\fbtobc$
and for extreme choices of the $\bc$ system parameters.
        
{\renewcommand{\arraystretch}{1.5}
\begin{table}                    
\begin{center}
\vspace {0.5cm}
\begin{tabular}{|c||c|c|c|} 
\hline 
  & \multicolumn{3}{c|}{$(\tan\beta/M_H)_{\rm min}$ (GeV$^{-1}$) } \\
  \hline                    
  $\fbtobc$ &            low & central & high  \\ \hline                   
 $2 \times 10^{-4}$  &   0.37 &       0.35 & 0.33 \\
 $1 \times 10^{-3}$  &   0.33 &       0.30 & 0.27 \\  \hline
\end{tabular}                                               
\ccaption{}{Lower limits on the ratio $\tan\beta/M_H$ (GeV$^{-1}$)
obtained by taking into account the contribution of $\bc$ mesons to the L3
search for $b\to \tau\nu$ final states.}
\end{center}                            
\end{table}
}

\begin{figure}[t]
\begin{center}
\epsfig{file=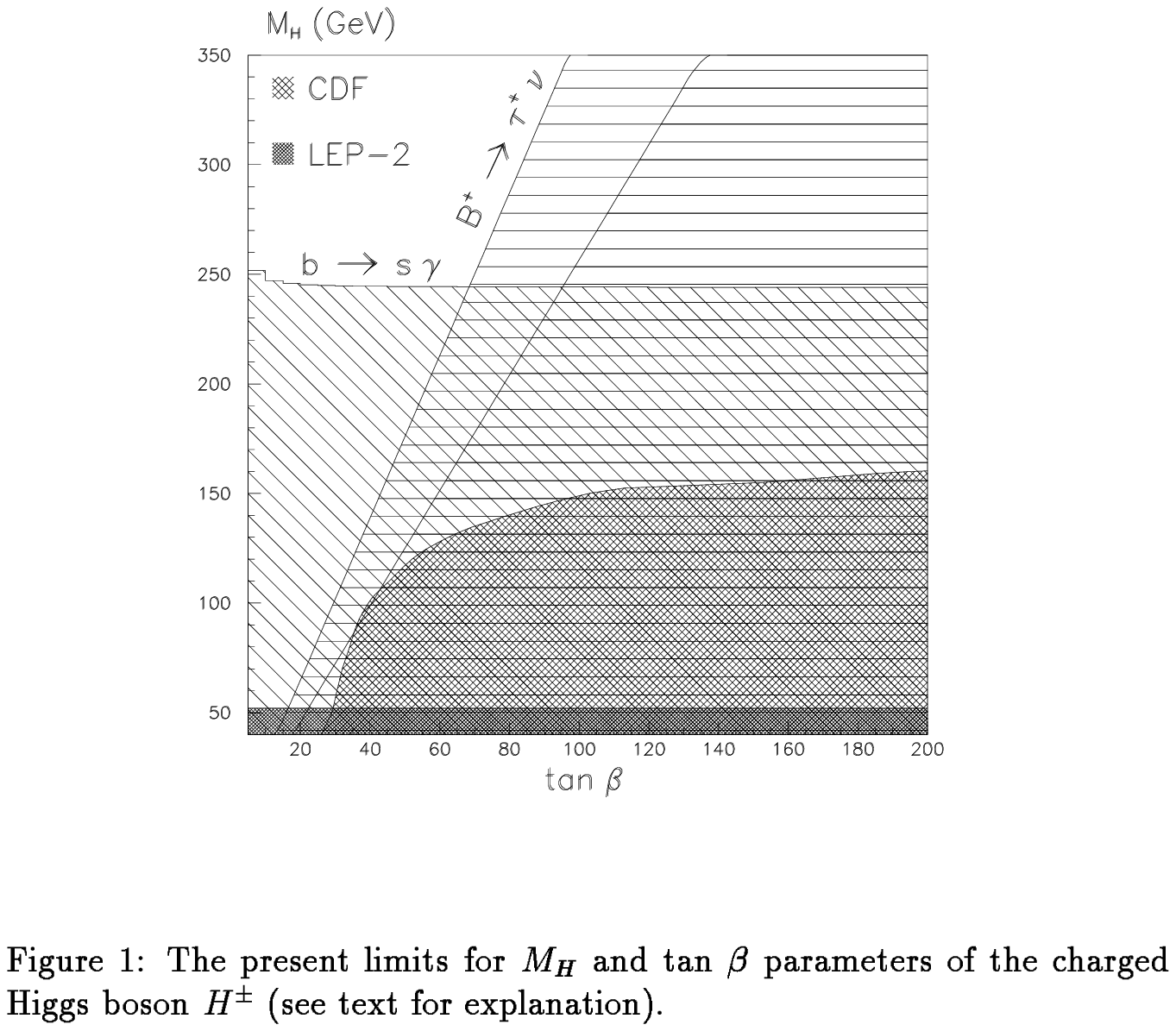,width=12cm,clip=}
\ccaption{}{Present limits for $M_H$ and tan~$\beta$. The two diagonal lines
next to the caption $B^+\to \tau\nu$ refer to the constraints derived from the
L3 limit. The right-most line is the L3 limit, the left-most line
corresponds to the 0.27 entry in table~2 of this work.}
\end{center}                                          
\end{figure}
In the Fig~1. we exhibit the present limits for $M_H$ and tan~$\beta$
parameters of the charged Higgs boson~$H^{\pm}$. In this figure we combine the
results from the direct searches of $H^{\pm}$ (CDF~Collaboration~\cite{cdf}
and DELPHI~Collaboration~\cite{leph}), from the measurements of the 
$b \to s \gamma$ transition (CLEO Collaboration~\cite {bsg}), and the results
from the search for $B \to \tau \nu$ decay (the right line represents original
L3~Collaboration~\cite{l3} result(\ref{l30})) while the left line 
corresponds to this article analysis). 
                                                                 
In summary, we discussed in this paper the possible effects of $\bc$ production
and leptonic decay on current $b\to \tau\nu$  searches at LEP. We showed that
the $\bc\to\tau\nu$ decay can produce a number of $b\to \tau\nu$ events as
large as the $\bu\to\tau\nu$ process. The uncertainties in the production and
decay properties of the $\bc$ meson are such that no precision measurement of
either $\fbu$ or $\vub$ can therefore  be expected from LEP, or similar
high-statistics measurements done at machines producing $b$ quarks well above
threshold. These problems will clearly disappear at the forthcoming
$B$-factories, where no contribution from $\bc$ mesons is possible.
We showed that the limits on possible new physics obtained from
the LEP measurements, e.g. leptonic decays mediated by charged Higgses, can
however be significantly improved when the effect of $\bc$ production and decay
is accounted for. These improved limits will become more compelling once the
leading uncertainties in the determination of the properties of the $\bc$
system will be improved. For example the detection of $\bc$ candidates in other
channels, such as $\bc\to \psi\ell\nu$, should allow to fix the main unknown,
namely the value of $\fbtobc$.

\noindent {\bf Acknowledgements}                                 
\noindent We are grateful to A. Falk, M. Neubert,
V.F.~Obraztsov and A.A.~Sokolov for useful 
discussions.                                                        
                                                     
\end{document}